\documentclass[usenatbib,useAMS]{mn2e}
\usepackage{times}
\usepackage{epsfig}

\begin{document}
\title[Illuminating Hot Jupiters in caustic crossing]{Illuminating Hot Jupiters in caustic crossing}

\author [Sedighe Sajadian and Sohrab Rahvar]
{Sedighe Sajadian and Sohrab Rahvar \thanks{rahvar@sharif.edu} \\
Department of Physics, Sharif University of Technology,\\
P.O.Box 11365--9161, Tehran, Iran}

\maketitle

\begin{abstract}
In recent years a large number of Hot Jupiters orbiting in a very
close orbit around the parent stars have been explored with the
transit and doppler effect methods. Here in this work we study the
gravitational microlensing effect of a binary lens on a parent star
with a Hot Jupiter revolving around it. Caustic crossing of the
planet makes enhancements on the light curve of the parent star in
which the signature of the planet can be detected by high precision
photometric observations. We use the inverse ray shooting method with
tree code algorithm to generate the combined light curve of the
parent star and the planet. In order to investigate the probability
of observing the planet signal, we do a Monte-Carlo simulation and
obtain the observational optical depth of $\tau \sim 10^{-8}$. We show that
about ten years observations of Galactic Bulge with a network of telescopes will
enable us detecting about ten Hot Jupiter with this method. Finally we show that the
observation of the microlensing event in infra-red band will
increase the probability for detection of the
exo-planets.
\end{abstract}

\section{Introduction}
Gravitational microlensing as one of the applications of general
relativity is proposed by Paczy\'nski (1986) for detecting the dark
compact halo objects so-called MACHOs in the Galactic halo. While
not enough MACHOs have been detected in the halo \cite{mil}, however
the microlensing technic has been used as an astrophysical tool for
studying the atmosphere of the stars and exploring the exo-planets.
In the standard method for exploring planets with the microlensing,
a star with the companion planets can play the role of lens
and produce caustic lines where crossing the caustics by the source
star produces a high magnification on the light curve
\cite{pac91}. In this case in addition to the standard microlensing
light curve we can detect a short duration spark due to the caustic
crossing, formed by the planet. A precise photometry
of the event is essential to find out this short duration signature
of the planet. The advantage of this technic compare to the other
methods of the exo-planet detection is that it is sensitive to the
observation of earth mass planets \cite{bal05} and also those
planets located beyond the snow line \cite{gould01}. There is also
other methods in gravitational microlensing such as planetary
microlensing signals from the orbital motion of the source star
around the common barycenter of source star--planet system
\cite{rahvar}.

In addition to the mentioned methods, Graff \& Gaudi (2000) proposed
caustic crossing of a close-in Jupiter size planet, produced by a
binary lens. In this case the planet's light is magnified so much
that it can be detected by a 10-m class telescope. Here we extend
this work looking to the details of the light curves and study
the most favorite pass band for this observation. Since in a
close-in Jupiter, the thermal emission due to the high temperature
of the planet is more significant than the reflected light from the
parent star, the observations in the
Infra-red pass band is more favorable than the visual pass band. We
also do a Monte-Carlo
simulation with a given observational strategy to obtain the number
of observable events in terms of the parameters of the planet and
the parent star. We emphasize that while the observations of the hot Jupiters
is simpler in nearby stars via the eclipsing and doppler methods, the microlensing
method can detect distant systems and enable us to compare the statistics of the
hot Jupiters with the nearby observations. One of the interesting features of the light curve
for the planet caustic crossing is that the planet can cross the
caustic more than that of parent star, as it traces effectively a
longer path due to the revolving motion around the parent star.

The paper is organized as follows. In Section \ref{lightcurve} we
will introduce the caustic crossing of the parent star and planet
system and generate
light curve with inverse ray shooting technic, introducing a new
development in tree-code algorithm. In section \ref{char} we study
the characteristics of the light curve in terms of the orbital
parameters of the planet and the parent star. In section \ref{mc} we
explain our Monte-Carlo simulation for estimating the probability
of illuminating Hot Jupiters with this method. In section \ref{conc}
we give the conclusions.

\section{Microlensing Light Curve}
\label{lightcurve} A binary system deflects the light ray with more
complicated way than a single lens. Let us represents $\xi$ as the position
of the image in the lens plane and $\eta$ the position of the source
in the source plane, the geometrical relation between
these parameters is given by the lens equation as follows \cite{sw86}:

\begin{eqnarray}
\label{lense}
\underline{\eta}=\frac{D_{s}}{D_{l}}\underline{\xi}-D_{ls}\underline{\alpha}(\underline{\xi}),
\end{eqnarray}
where underline represents the vector, $\underline{\alpha}$ is the
overall deflection angle due to a double lens and $D_{s}$, $D_{l}$
and $D_{ls}$ are the distance of the source and lens from the
observer and distance between the lens and source, respectively. The
deflection angle for the light ray is given as follows
\begin{eqnarray}
\underline{\alpha}(\underline{\xi})=\frac{4GM_{1}}{c^{2}}
\frac{\underline{\xi}-\underline{\varrho}_{1}}{|\underline{\xi}-\underline{\varrho}_{1}|^{2}}
+\frac{4GM_{2}}{c^{2}}
\frac{\underline{\xi}-\underline{\varrho}_{2}}{|\underline{\xi}-\underline{\varrho}_{2}|^{2}},
\end{eqnarray}
where $\underline{\varrho}_{1}$ and $\underline{\varrho}_{2}$ are
the positions of the binary system in the lens plane and $M_{1}$ and
$M_{2}$ are mass of the lenses. Equation (\ref{lense}) as the
lens equation is a fifth order equation and in general the solution
is not trivial. One of the possible solution for solving the lens
equation is the inverse ray shooting method \cite{kay86,sw87}. In
this method we follow the position of the light ray that shoots from
the observer to the lens plane, knowing the position of the lenses
we can calculate the deflection angle and substituting in the lens
equation results in the position of the source. We pixelize the
source and the lens plane and for each light ray
passing from the lens plane and hitting the source plane, we count
the number of hits inside each pixel. These numbers identify the
magnification pattern in the source plane. We use tree code method
as we describe it later, for generating the image and the
magnification of the source star.

In order to simplify our calculation we take the dimensionless
parameters in the lens equation.  Let us define the overall Einstein
radius as follows:
\begin{eqnarray}
R_{E}=\sqrt{\frac{4G(M_1 + M_2)}{c^{2}} \frac{D_{l}D_{ls}}{D_{s}}}.
\end{eqnarray}
We normalize the lens equation to this length scale, which results in:
\begin{eqnarray}
\underline{x}=\underline{r}-\underline{\alpha}( \underline{r} ),
\end{eqnarray}
where $\underline{x}={D_{l}}/{D_{s}}\times \underline{\eta}/{R_{E}}$
, $\underline{r}=\underline{\xi}/R_{E}$ and the deflection angle
is given by:
\begin{equation}
\underline{\alpha}(r) = \mu_1
\frac{\underline{r}-\underline{r}_{1}}{|\underline{r}-\underline{r}_{1}|^2}
+
\mu_2\frac{\underline{r}-\underline{r}_{2}}{|\underline{r}-\underline{r}_{2}|^2}
\end{equation}
and $\mu_i = M_i/(M_1+M_2)$ ,
$\underline{r}_{i} =
\underline{\varrho}_{i}/ R_{E}$.

We take a straight line for the path of the center of mass of the parent
star and planet at the lens plane. Taking the mass of parent star
larger than the mass of planet, the parent star follows
approximately a straight line as follows:
\begin{equation}
\underline{{u}_{s}}=(-u_{0}\sin\alpha+\frac{t-t_{0}}{t_{E}}\cos\alpha)\hat{i}+
(u_{0}\cos\alpha+\frac{t-t_{0}}{t_{E}}\sin\alpha)\hat{j},
\end{equation}
where $u_{0}$ is defined  as the minimum impact parameter of source
star from the center of the cartesian coordinate system, normalized
to the Einstein radius, $t_{0}$ is time of impact parameter, and
$\alpha$ is angle between $x$-axis and the trajectory of the source
star. The rotation of the planet around the parent star makes an
cycloid like pattern on the source plane which is given by:
\begin{eqnarray}
\label{pt} \underline{{u}_{p}}&=&\underline{{u}_{s}}
+\tilde{a}[\cos(\omega t+\varphi)\cos(\alpha+\beta) \nonumber \\
&-&\cos(\delta)\sin(\omega t+\varphi)\sin(\alpha+\beta)]\hat{i}\nonumber\\
&+&\tilde{a}[\cos(\omega t+\varphi)\sin(\alpha+\beta)\nonumber\\
&+&\cos(\delta)\sin(\omega t+\varphi)\cos(\alpha+\beta)]\hat{j},
\end{eqnarray}
where $\tilde{a}$ is the projection of the planet orbit on the lens plane normalized to the
Einstein radius, $\omega$ is the angular velocity of planet around the source
star which can be obtained from the Kepler's third law:
\begin{equation}
\omega = \sqrt{\frac{G(M_p+M_\star)}{a^{3}}},
\end{equation}
and $a$ is the orbital radius of the planet, $\delta$ is the
deviation of the normal vector to the orbital plane of the planet
from our line of sight. For the Hot Jupiters
due to the tidal interaction of the
planet and the parent star we set the eccentricity to zero and one
angle is sufficient for describing the orbital
plane deviation,
$\beta$ is angle between the trajectory of the source star and the
projected semi-major axis of the planet and $\varphi$ is initial
phase of the planet.

In order to generate the light curve, we need the relative velocity
of the binary lens with the parent star and companion planet.
%If we take the source star at the Galactic Bulge and the lens stars at
%the disk or the bulge,
The relative transverse velocity of the
source-observer line of sight with respect to the lens at the lens
plane is given by \cite{kay86}:
\begin{equation}
\underline{v_{rel}}=x\underline{v_{s}} - \underline{v_l} - \underline{v_o}(1-x),
\end{equation}
where $\underline{v_l}$, $\underline{v_s}$ and $\underline{v_o}$ are
the two dimensional transverse velocities of the center of mass of
the lens system, source and observer with respect to the line of
sight respectively and $x=D_l/D_s$ is the ratio of the distance of
the lens to the source. The velocity of the observer is
obtained from the local measurements of the solar system in the
Galactic frame. The velocity of the stars in the Bulge
is given by the dispersion velocity in this
structure and the velocity of the stars in the disk is
obtained from the combination of the dispersion and
global velocities of the stars \cite{bin}.

Since the lens is a binary system, that will rotate around the
center of mass during the microlensing of the parent star and the
companion planet. For simplify in our calculation we fix the position
of the binary system and obtain the relative motion of the source
objects with respect to the rotating binary system. The situation is similar to the studying
of the motion of an object in rotating non-intertidal reference
frame in the classical mechanics. In the reference frame of the
binary system we do the following coordinate transformation for the
position of the source object
\begin{equation}
{\cal R}(\Omega,\psi) = \left(\begin{array}{cc}
\cos\Omega t& -\sin\Omega t \\
\sin\Omega t & \cos\Omega t
\end{array}
\right)\,
\label{eq:rotmatrix}
\end{equation}
where $\Omega$ is the angular velocity of the binary lens,which is
given by
\begin{equation}
\Omega = \sqrt{\frac{G(M_1+M_2)}{d_\bot^3}\cos\psi^3},
\end{equation}
and $d_\bot$ is the apparent separation of the two lenses from each
other and $\psi$ is the deviation of the normal vector to the binary
plane from the line of sight.

\begin{figure}
\begin{center}
\psfig{file=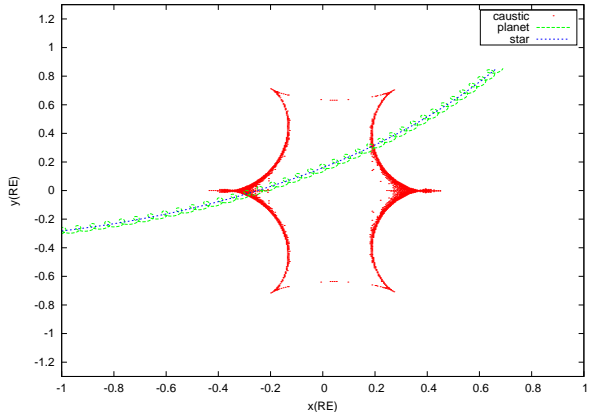,angle=0,width=8.cm,clip=} \caption{ The
trajectory of the parent star with the companion planet in the source
plane is compared with the caustic lines of a binary lens. The scale
in this figure is given in terms of overall Einstein radius of a
binary lens.} \label{fig1}
\end{center}
\end{figure}

For generating the light curve we should note that there are two
source objects, the parent star and the planet. The total flux
receiving by the observer then is the accumulation of the magnified
flux of each component \cite{hg97,gh92}:
\begin{equation}
\label{mag} A_{total} = \frac{F_{\star} A_{\star}+F_{p}
A_{p}}{F_{\star}+F_{p}},
\end{equation}
where the $F_{\star}$ and $F_p$ are the intrinsic flux of the parent
star and the planet and $A_\star$ and $A_p$ are the corresponding
magnifications. Let us define the ratio of the intrinsic flux of the
planet to the flux of the parent star by $\varepsilon = F_p/F_\star$
which is much smaller than one (i.e. $\varepsilon\ll 1$). The total
magnification can be written as:
\begin{equation}
A_{total} = A_{\star}+\varepsilon A_{p}. \label{maga}
\end{equation}
For a typical case of the binary lens and a main sequence parent star
with an Hot Jupiter companion, the trajectory of the source in
the source plane with the corresponding caustic lines is shown in
Figure (\ref{fig1}).
\begin{figure}
\begin{center}
\psfig{file=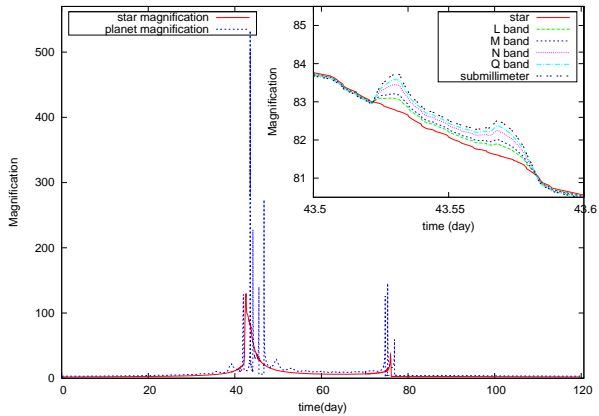,angle=0,width=8.cm,clip=} \caption{ The
magnification of parent star (solid-line) and the companion planet
(dashed-line) in terms of time for the configuration given in Figure
(\ref{fig1}), for various pass bands. In the top-right hand side of
figure, the overall magnification is given in terms of time around
the first star's peak surrounded by the two peak of the
planet.}
\label{fig2}
\end{center}
\end{figure}
\begin{figure}
\begin{center}
\psfig{file=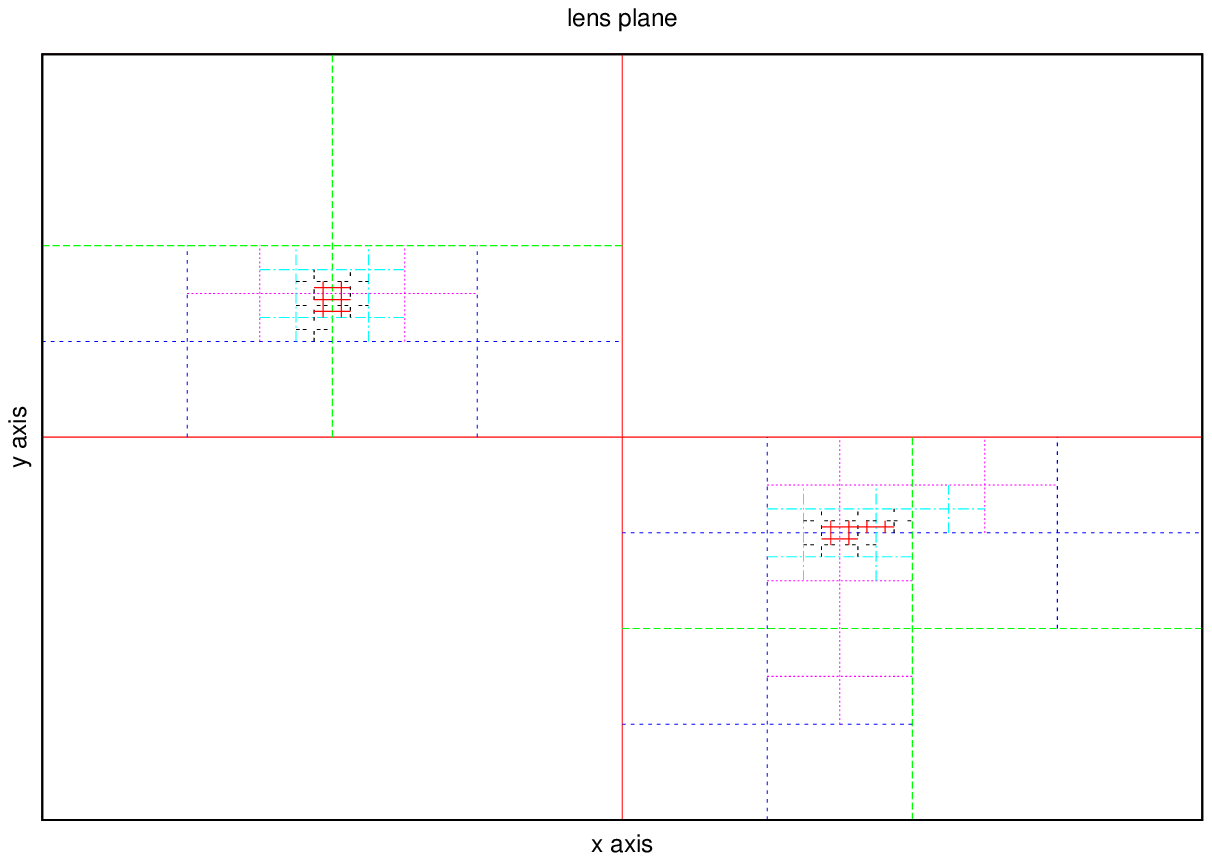,angle=0,width=8.cm,clip=}
\psfig{file=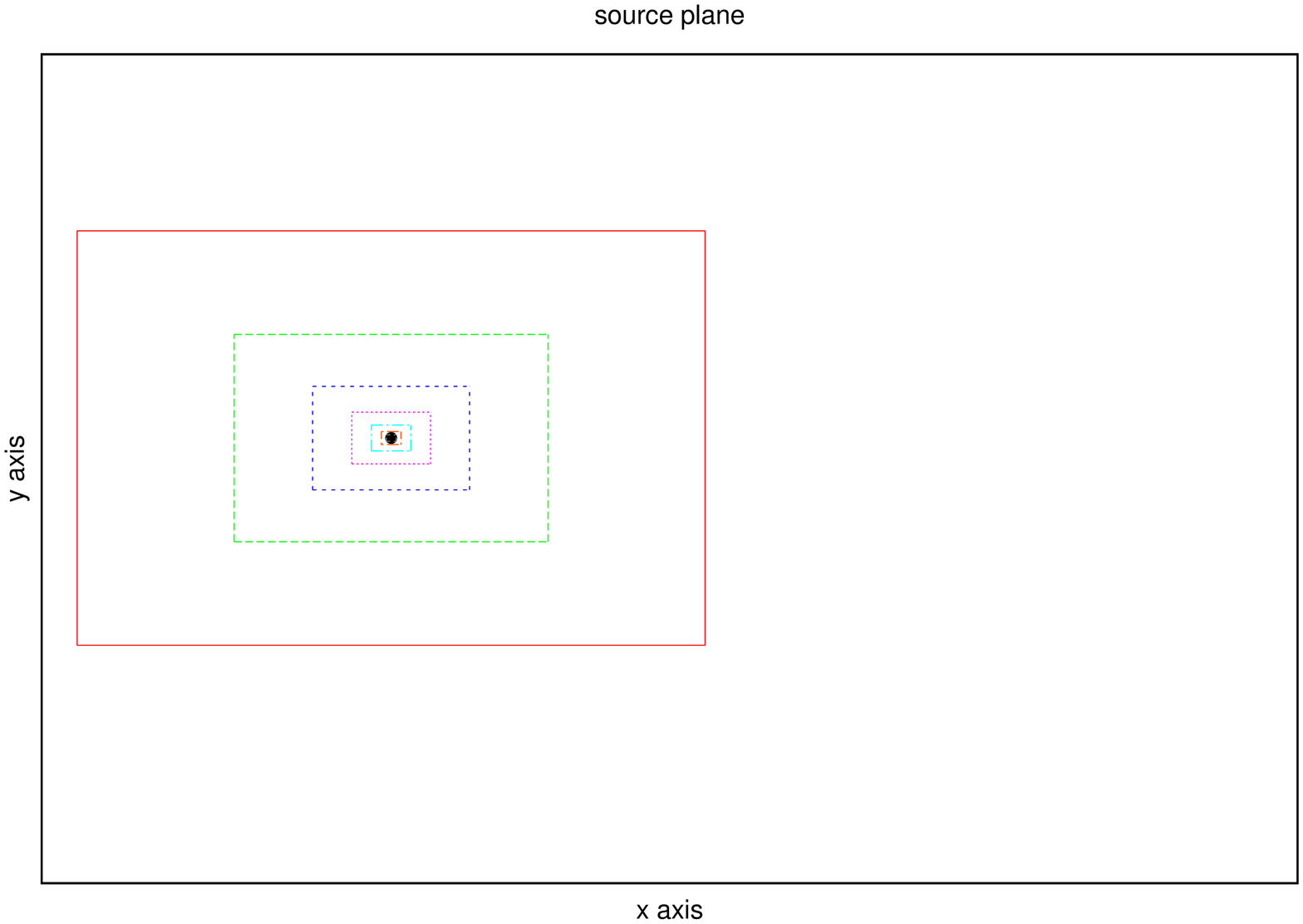,angle=-90,width=8.cm,clip=}
\caption{Pixelizing process is decipted in this figure. The Upper
panel represents the lens plane and the lower panel is the source
plane. We take the same size pixel for these two planes during the
dividing process. For generating the light curves as shown in Fig. (\ref{fig2}), we
cover the parent star with $5\times5$ gride.} \label{fig3}
\end{center}
\end{figure}
One of the interesting features in the caustic crossing is that
while the parent star crosses the caustic lines for two times, the
planet due to the cyclic motion approach to the caustic lines more
than that of the parent star. The corresponding light curve of the
source and planet is shown in Figure (\ref{fig2}). Even during the
caustic crossing of the parent star in the first peak, the planet
crosses twice the caustic lines. In order to distinguish the caustic
crossing of the planet in the first peak, we zoom in around the
first peak of the parent star which is given in the right corner of
the Figure (\ref{fig2}). We note that the re-crossing of the caustic by
the planet takes place with an interval time of about one hour and in
order to detect this signal we need a sampling rate less than this time scale.

In the rest of this section we introduce the details in generation
of the light curve in the numerical calculation. We should note
that the tree code method used in our calculation is different than
that introduced by Barnes \& Hut (1986) and Wambsganss (1999). In
their method the lenses in the lens plane are treated corresponding
to their distances to the light ray. In that approach the process of
making cells on the lens plane depends on the distance of the lenses
to the light ray and their distribution. For the weak gravitational
regime where the deflection angle is obtained from the superposition
of each lenses, the closer lenes enter to the calculation of the
deflection angle with more accuracy than the distant ones. This
method can be used for a large number of lenses as the passage of a
Quasar light ray through a galaxy.

In our tree code method we divide the lens plane into the cells
and choose a same size pixel to surround the source at
the source plane. The procedure starts with dividing the lens plane
into four parts and in the source plane we choosing a pixel with the
same size to surround the source object where the source is located at the
center of the cell. We shoot enough number of light rays to each
cell in the lens plane to cover the cell's area. Using the lens
equation, if at least one of the rays collides to the source cell,
we accept that cell for the next dividing step otherwise we remove
this cell from the lens plane. We continue this dividing procedure
up to the stage that we have enough pixels to cover the source
object, having reasonable resolution. As much as smaller cells
provides better resolution and less statistical fluctuations in the
magnification calculation. Since this pixelizing is time
consumer, specially in the Monte-Carlo simulation,
we stop the dividing procedure up to a desire resolution.
For instance for the parent star, we cover the source with a
$5\times5$ gride. Figure (\ref{fig3}) depicts the pixelizing
procedure in this algorithm. In the final stage of pixelizing, the
remained number of cells in the lens plane to that of cells in the
source plane that covers the source star provides the magnification
due to the gravitational lensing. Also the remained pixels in the
lens plane provides the shape of the images. We should note that due
to the smaller size of the planet with respect to the source star,
the magnification of the planet dose not suffer as much as the
parent star from the finite-size effect. The result of planet
caustic crossing will be like a flashing, let's call it "planet
flashing". While this flashing in term of magnification is much
larger than that of magnification of the source star (see Figure
\ref{fig2}), the intrinsic flux of the planet relative to the flux
of the source star should be sufficient to be observed by a
telescope. In the next section we will discuss this issue in details.

\section{Characteristics of the light curve}
\label{char} In this section we study the details of light curve
looking to the physical specifications of the source star and the
companion planet as well as the binary lens. One of the main
factors in the observability of the planet by this method is the
ratio of the planet's flux as the signal to the flux of parent star
as the background. Using equation (\ref{maga}) we can describe the
signal to the background in terms of the total and parent star
magnifications:
\begin{equation}
\frac{\delta F}{F} = \frac{A_{total}}{A_\star} - 1.
\end{equation}
The overall flux receiving from a planet to the observer contains
the radiation of the thermal energy due to the intrinsic temperature
of the planet and the reflection of the parent star's light.
Assuming the thermal emission of the planet as a black body
radiation \cite{lop07}, the temperature of the planet can be
calculated by taking into account the absorption of the radiation
receiving from the parent star and reradiating it through the
Boltzman law. Hence the planet's temperature is given by:
\begin{equation}
\label{tpts}
T_{p} = T_{\star}(\frac{R_{\star}}{a})^{1/2}[f(1-A_{B})]^{1/4},
\end{equation}
where $R_{\star}$ and $T_\star$ are the radius and the temperature of
the parent star, $A_B$ is the Bond albedo and $f$
describes the fraction of reradiating energy which is absorbed by
the planet. Recent studies on the difference between the day and
night luminosity of the planets through eclipsing show that $f$ can
change between $1/4$ and $2/3$ \cite{har06,knu07}. $f=2/3$
represents a Hot Jupiter with low advection where in this case there
is a big different between the temperature of the day and night on
the planet and $f=1/4$ represents a planet with high redistribution
of energy.

The thermal flux of the planet is given by the Planck's law as
\begin{equation}
\label{sbeq}
I(\nu,T_p)= \frac{2h\nu^3}{c^2} (e^{\frac{h\nu}{k T_{p}}}-1)^{-1},
\end{equation}
where $I(\nu,T_p)$ represents the emitted power per unit area of
emitting surface, per unit solid angle and
per unit frequency. Integrating over the frequency and over
an half-sphere results in the Stefan-Bolttzmann law as follows:
\begin{equation}
F_p ^{(th)}= (\frac{R_p}{D_s})^2\sigma T_p^4,
\end{equation}
where $R_p$ is the radius of the planet. On the other hand the flux
of the planet due to the reflection of parent star's light is given
by
\begin{equation}
F_{p}^{(ref)} = {F_{\star}}A_g (\frac{R_{p}}{a})^{2}g(\Phi),
\label{ref}
\end{equation}
where $A_g$ is the geometrical albedo, assuming the Lambert's law,
$A_g=2/3A_B$, $\Phi$ is the phase of the planet and $g(\Phi)$ is
a function of the planet phase indicates a fraction of the lighted
area of planet in front of the observer. Now the ratio of overall
planet's flux which is composed by the thermal and reflation terms
to the star's flux is given by:
\begin{equation}
 \frac{F_p}{F_\star} = g(\Phi)[A_g(\frac{R_p}{a})^2 +
(\frac{R_p}{R_\star})^2(\frac{T_p}{T_\star})^4],
\end{equation}
where eliminating the planet's temperature from equation
(\ref{tpts}), the flux radio obtains in the simpler form of
\begin{equation}
\label{ow}\frac{F_p}{F_\star} = (\frac{R_p}{a})^2[A_g +
f(1-A_B)]g(\Phi).
\end{equation}
The important point in this equation is that it is independent of
the temperature and the radius of the source star and it depends only
on the chemical composition of the planet atmosphere which changes
$A_B$, the size and the distance of the planet from the parent star.
During the gravitational lensing we multiply the corresponding
magnifications of the planet and the parent star to the intrinsic
flux of the source objects and obtain the overall receiving flux.

In practice we do not integrate over all the wavelengths, as the
detector of the telescope may sensitive to an specific pass band. To
find the optimal wavelength for the detection, we maximize
the ratio of the intensity of the planet to that of the the parent
star, i.e. $\delta I(\nu)/I(\nu)$. Using equation (\ref{sbeq}) and (\ref{ref}),
assuming a constant reflection index for all the wavelengths (i.e.
$dA_g/d\nu \simeq const$), this ratio obtain as:
\begin{equation}
\label{fr} \frac{\delta I(\nu)}{I(\nu)}=
[(\frac{R_p}{a})^2\frac{dA_g}{d\nu} +
\frac{e^{h\nu/kT_\star}-1}{e^{h\nu/kT_p}-1}(\frac{R_p}{R_\star})^2]g(\Phi).
\label{ratio}
\end{equation}
\begin{figure}
\begin{center} \psfig{file=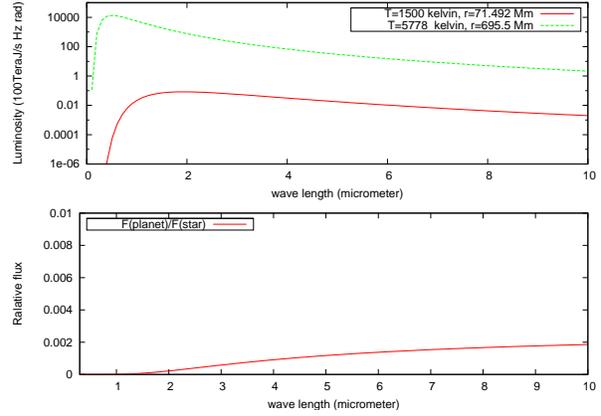,angle=0,width=8.cm,clip=}
\caption{Comparison of luminosity between the parent star and the
planet. At the upper panel, solid line represents the luminosity of the
planet with $T_p=1500~K$ and the dashed line is given for the parent star
with $T_\star =5778~K$. The radius of the planet and the source star
is given in Mega-meter. The lower panel shows the ratio of the
luminosity for the planet to the parent star.} \label{fratio}
\end{center}
\end{figure}
We should note that the reflection flux is mainly important in
the optical bands, where the thermal flux dominates
in the infra-red and sub-millimeter wavelengths.

\begin{figure}
\begin{center}
\psfig{file=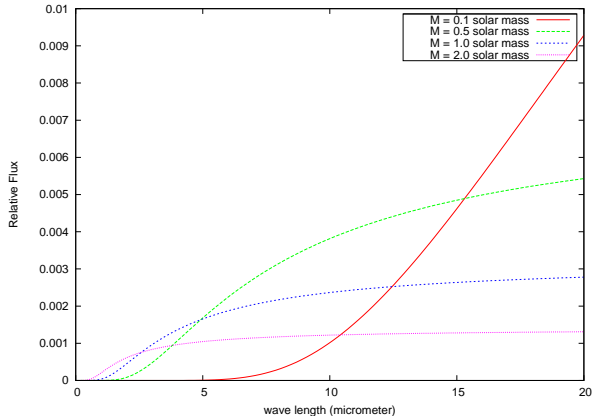,angle=0,width=8.cm,clip=} \caption{ The ratio
of the planet flux to the parent star as a function of wavelength
for four different mass of the parent star. The size and the
distance of the planet is fixed to the Jupiter mass with an orbital
radius of $0.05 Au$. } \label{mass}
\end{center}
\end{figure}

Hot Jupiters have small albedo of $A_{g}\leq0.2$ \cite{ro08}, so
most of the radiation of the parent star is absorbed by the planet's
atmosphere and is reradiated in the infra-red band. The result is a
small share of luminosity of the planet in the reflected flux. The
exception in the Hot Jupiters happens in the cases with the period
of motion smaller than three days. This class of Hot Jupiters
so-called Very Hot Jupiters has a surface temperature larger than
$2500~ K$ in which they reradiate considerable amount of thermal
flux in the optical pass band. However, the peak of the spectrum is
in infra-red band \cite{lop07}. Out of this exception, the most Hot
Jupiters have a considerable flux in infra-red due to the thermal
emission \cite{ds09}. We plot equation (\ref{fr}) in Figure
(\ref{fratio}) as an evidence that in the infra-red pass band, we
have the most contrast between the flux of the planet and the parent
star. Here we compare the planet luminosity to the parent star
luminosity as a function of the wavelength. In order to see the
sensitivity of this fraction to the mass of the parent star, we plot
this function for four masses of the parent star. Choosing the
parent star as a main sequence star, the radius of the star can be
eliminate in favor of the mass, using
\begin{equation}
\label{ms}
 R_{\star} = M_{\star}^{0.8},
\end{equation}
where the corresponding parameters are normalized to that of sun's
values. Figure (\ref{mass}) shows that M-dwarf stars are more
favorable for the Hot Jupiter detection in the wavelengths longer
than $15\mu~m$. However, we should note that for the M-dwarf star, the
abundance of the Hot Neptunes is more than the Hot Jupiters
\cite{{Ida05}}.

\section{monte-carlo simulation}
\label{mc} In this section we do a Monte-Carlo simulation to obtain
quantitatively the sensitivity of the planet detection to the
parameters of the binary lens and the planetary system. Finally we
provide the probability for detecting desired events.

First of all we need to generate the parameters of the lens and the
source according to the physical distribution of the parameters. We
divide the parameter space into the lens and the source parameters
for better analyzing. For the lens parameters we take $q =
M_{1}/M_{2}$, the ratio of the masses in the binary system to change
uniformly in the range of $q\in[0,1]$, the mass of one of the lenses
is taken from the Salpeter mass function \cite{sal95} in the range
of $M_{1}\in[0.1,3]M_\odot$ and the distance between the lenses is
taken uniformly in the range of $d\in[0.1,5]~A.U.$. The location of
the lens from the observer is calculated from the probability
function of microlensing detection $d\Gamma/dx\propto
\rho(x)\sqrt{x(1-x)}$ where $x$ changes in the range of $x\in[0,1]$.
The velocity of the lens is taken by the combination of the global
\cite{ero09} and the dispersion velocity \cite{bin} of the disk and
bulge. We take a thin disk for modeling the disk and take our line
of sight towards the Galactic bulge with the latitude angle in the
range of $b\in[1,2]$.

For the source objects, the corresponding angles in the trajectory
of the source system, $\alpha$, $\phi$ and $\beta$ in Equation
(\ref{pt}) is taken uniformly. The minimum impact parameter is in
the range of $u_0 \in[0,1]$. we take the mass of the parent star
from the Salpeter mass function in the range of $M_\star\in[0.1,3]
M_\odot$. Since the parent star is assumed to belong to the main
sequence, the radius of the star is taken from equation (\ref{ms}).
On the other hand the luminosity of the parent star can be obtained
from \cite{edd26}:
\begin{equation} L_\star =L_{\odot}(\frac{M_\star}{M_{\odot}})^{3.5}.
\end{equation}

For the parameters of the planet, the inclination angle $\delta$ of
the planetary plane to the line of sight is taken uniformly in the
range of $\delta\in[-\pi/2,\pi/2]$.
%Due to the energy loss of the
%Hot Jupiters during the tidal interaction with the parent star we
%take circular orbit for the trajectory of the planet.
We take the
mass of the planet in the range of $M_{p}\in[0.1,10] M_{J}$ and
distance of the planet from the parent star in the range of
$a\in[0.01,0.1]~A.U$. For the close-in Hot Jupiters at the distance
less than $\sim 0.05~A.U.$, which are in our concern, the radius not
only is a function of the planet's mass, it also depends on the
distance of the planet from the parent star \cite{fort07}. Close-in
planets are heated by the parent star and their atmosphere inflates.
For the hot planets, in contrast to the conventional relation
between the mass and the size, the small mass planets, inflate more
than the massive ones. We take the mass-radius-distance relation of
Hot planets from Fortney {\it et al.} (2007).

Before performing  Monte-Carlo simulation to count the number of
events with the desired signal to the background flux, we do a first
order estimation, just counting geometrically the number of planets
that their trajectories cross the caustic lines. In our simulation
we follow the path of the planet in the magnification pattern that
has already been generated by the inverse-ray shooting and assign
the magnification of the source object along its trajectory. We
calculate the deviation of the flux along the path to identify any
sharp peak in the light curve. This signature indicates a caustic
crossing. Our simulation shows that in microlensing events with
binary lens and for the condition of $u_0<1$, almost $36\%$ of
planets can cross the caustic lines. We obtain almost the same
amount of the caustic crossing for the parent stars. We call this
fraction of events with the caustic crossing as the geometrical
criterion for the Hot Jupiter detection.

In reality we should measure the flux of source star and the
enhancements due to the Hot Jupiter on the background light curve.
First we assume a telescope with $1\%$ photometric precision. We can
change this criterion according to the size of telescope and a
limiting magnitude on the brightness of the source star. With new
technics as defocusing of the telescope, we can achieve dispersions
in the order of $5~mmag$ with a medium size telescope where in
terms of the relative flux dispersion to the background, it
is in the order of $10^{-4}$ \cite{sw09}.

In our Monte-Carlo simulation we look to the maximum magnification
of the planet when it crosses the caustic line and obtain the flux
ratio of the planet to the source star at that point. Here we do not
take into account the sampling rate of the observations of the
event. A typical duration for the magnification of the planet during
the caustic crossing is in the order of one hour and we assume to
have a network of the telescopes to cover the event. For the
photometic precision of $1\%$, in Table (\ref{table1}) we show the
detection efficiency for various pass bands and atmospheric models
of the Hot Jupiters. Longer wavelengths are more favorable for
detection of the planet than the shorter ones. Also planets with
high reradiating property (i.e. $f=2/3$ in this simulation), means
less advection in the atmosphere are more favorable for the
detection.

\begin{table*}
\begin{center}
\begin{tabular}{|c|c|c|c|c|c|c|c|c|c|c|c|c|c|}
\hline \hline
& Band : &          & $Optical$ & $R$  & $J$ & $H$  & $K$  & $L$  & $M$  & $N$  & $Q$  & $Submilimeter$ & Overall wave length \\
& Wave length$(micrometer)$ & & 0.55 & 0.825 & 1.25 & 1.65 & 2.2&3.45&4.7 & 10 & 20 & 450 &   \\
\hline\\

& $A_{B}=0$ ; $f=\frac{1}{4}$ &
&0.01 & 0.04&0.13& 0.23&0.36&0.67&0.94&2.62&5.33&10.74&0.01 \\

&$A_{B}=0$ ; $f=\frac{2}{3}$ &
&0.09& 0.21 & 0.39 & 0.57 &0.85&1.50&2.38&5.85&9.33&14.35&0.11 \\

&$A_{B}=0.3$ ; $f=\frac{1}{4}$ &
&0.03& 0.06 & 0.13 & 0.21 &0.32&0.57&0.80&2.06&4.45&9.81&0.03 \\

&$A_{B}=0.3$ ; $f=\frac{2}{3}$ &
&0.08& 0.18 & 0.34 & 0.49 &0.70&1.19&1.80&4.72&7.90&13.18&0.11 \\

&$A_{B}=0.5$ ; $f=\frac{1}{4}$ &
&0.05& 0.07 & 0.13 & 0.20 &0.29&0.51&0.70&1.60&3.66&9.0&0.06 \\

&$A_{B}=0.5$ ; $f=\frac{2}{3}$ &
&0.09& 0.16 & 0.29 & 0.41 &0.58&0.99&1.40&3.77&6.69&12.08&0.11 \\

\hline
\end{tabular}
\label{table1}
\end{center}
\caption{Detection efficiency of Hot Jupiter for various
optical, infrared and far-infrared bass-bands. The first column
describes the atmospheric model of the planet, the second and the
third columns represents the efficiency for the optical and red
bands, the forth to the tenth columns shows the detection efficiency
for the infra-red and far infrared pass bands. The eleventh column
stands for the submilimeter and the last column shown the average
detection efficiency over all the wavelengths, weighted to the
spectrum of the planet.
\label{table1} }
\end{table*}

\begin{figure}
\begin{center}
\psfig{file=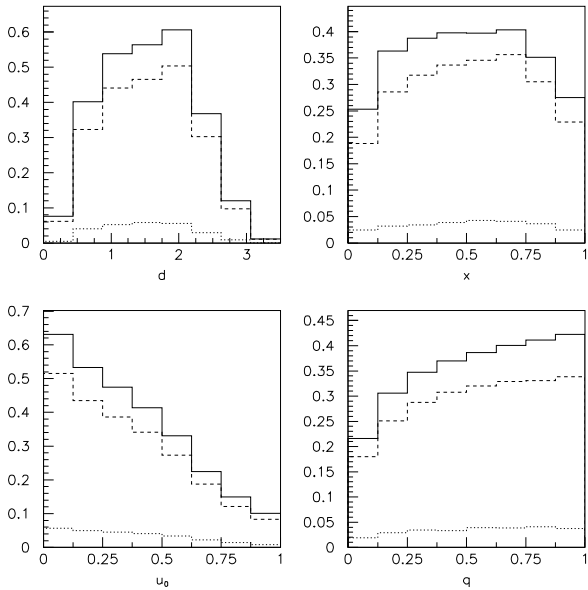,angle=0,width=9.cm,clip=} \caption{The
detection efficiency for different parameters of lens for three
different observational strategy with $10^{-2}$ (dotted-line),
$10^{-3}$ (dashed-line) and $10^{-4}$ (solid-line) photometric
precision.} \label{fig6}
\end{center}
\end{figure}

In order to see the sensitivity of the planet detection on the
parameters of the model, we plot the detection efficiency in terms
of the relevant parameters of the lens, source and planet in Figures
(\ref{fig6}) and (\ref{fig7}) for three cases of photometric
precision of $10^{-2}$, $10^{-3}$ and $10^{-4}$. We ignore the
irrelevant parameters that do not enter in the efficiency function.
The detection efficiency function in terms of the lens and the
source-planet parameters is given as follows:

\begin{figure}
\begin{center}
\psfig{file=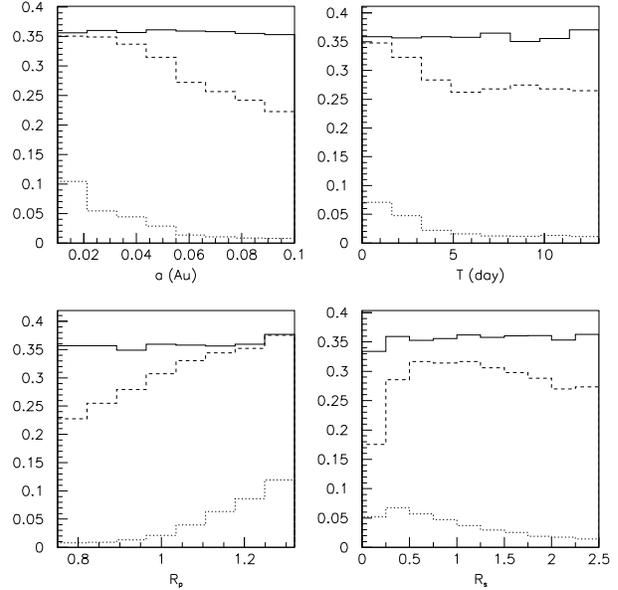,angle=0,width=9.cm,clip=} \caption{The
detection efficiency for different parameters of the planet and the
parent star for three different observational strategy with
$10^{-2}$ (dotted-line), $10^{-3}$ (dashed-line) and $10^{-4}$
(solid-line) photometric precision.} \label{fig7}
\end{center}
\end{figure}

(a) The distance between two lenses. Here the detection efficiency
raises with increasing the distance between the two lenses and after
a peak around $\sim 2R_E$, it decreases. The physical interpretation
of this feature is due to the topological configuration of the
caustic lines. It is shown in \cite{sw86} that for the case of $q=1$
and distance between the lenses in the range of
$[\frac{2}{\sqrt{8}},2]R_E$, the caustic lines are topologically
connected whereas beyond this range the caustic lines detach from
each other. Having continues caustic lines increases the probability
of caustic crossing both by the parent star and the planet.

(b) The second parameter is $x = D_l/D_s$, the relative distance of
the lens to the source star. In the detection efficiency diagram,
the efficiency reaches to the maximum value around $\sim 0.5$ where
the Einstein radius has the maximum size. A larger Einstein radius
results in a longer duration for the microlensing event and a higher
probability for the caustic crossing.

(c) The third parameter is the impact parameter. Decreasing the impact parameter
increase the detection efficiency. Smaller impact parameter from the center of
lens configuration increases the probability of the caustic crossing. On the other
hand statistically the impact factor may also affect on the magnification
of the light curve. We test this hypothesis amongst the simulated events showing
that the small impact parameters results in a higher signal to the background flux.

(d) The ratio of the lens masses, $q$. It seems that changing $q$
has a geometrical effect on the shape of the caustic lines,
where increasing it towards the symmetric shape, $q=1$  maximize
the detection efficiency.

\begin{table*}
\begin{center}
\begin{tabular}{|c|c|c|c|c|c|}
\hline
& $Photometric~Precision:$ &            & $10^{-2}$ & $10^{-3}$  & $10^{-4}$  \\
\hline\hline
& $\tau_p$ &                & $1.34\times10^{-9}$   & $1.34\times10^{-8}$    & $1.57\times10^{-8}$  \\
\hline
&$N_p$ &                              & 3 &  28 & 32  \\

\hline
\end{tabular}
\end{center}
\caption{The first line indicates the photometric precision, the
second line is the corresponding optical depth of the hot Jupiter
detection and the third line is the number of the planets that can
be detected by monitoring $10^7$ stars during $10$ years towards the
Galactic Bulge. \label{tab2}}
\end{table*}

(e) The semi-major axis of the planet, $a$, shown in Figure (\ref{fig7}). The Hot Jupiters reside
in the range of $[0.01,0.1]Au$. For the smaller $a$ there is larger
detection efficiency than the larger $a$. The dependence of the
detection efficiency to the semi-major axis results from the
reflection of the flux of the parent star from the planet,
proportional to inverse square of the distance. The other effect is
due to the intrinsic thermal luminosity of the planet, closer to the
parent star makes the temperature of planet higher.

(d) Period of motion of planet, T. The dependence of the efficiency
to the period is in the same way as the semi-major axis due to the
Kepler's law. Shorter duration for the period of the planets
resembles to the smaller semi-major axis.

(f) Radius of planet, $R_p$. The intrinsic flux of planet both in
reflection of the parent star's light and thermal emission is
proportional to square of radius of planet. Hence larger planets can
be detected easier than the smaller ones.

(g) Radius of source star, $R_s$. The effect of the radius of the
source star on the relative intensity of the planet to the parent star
can be seen in equation (\ref{ratio}). Increasing the radius of the
star causes decreasing this ratio.

Finally we come to conclude on the possibility of detection of the
hot Jupiters from this method. The comparison of the OGLE data with
the model constructed from the Hipparcos data indicates that about
$f_p \sim 0.5\times10^{-2}$ fraction of the stars have hot Jupiters
and very hot Jupiters \cite{gould06}. On the other hand the radial
velocity observations of the planets indicate that about $f_p \sim
10^{-2}$ fraction of the solar type stars have hot Jupiters
\cite{mar05}. From the Monte-Carlo simulation we obtain the average
detection efficiency $<\epsilon_p>$, for the planet detection with
three photometric precession of $10^{-2}$, $10^{-3}$ and $10^{-4}$
as $0.03$, $0.30$ and $ 0.35$. The optical depth for the planet
detection can be obtained by
\begin{equation}
\tau_{p} = <\epsilon_p> \times f_p \times \tau,
\end{equation}
where $\tau$ is the optical depth for the microlensing events
towards a given direction. For the direction of the Galactic Bulge
$\tau = 4.48\times 10^{-6}$ \cite{sum05}, hence the optical depth
for the planet detection is about $\tau_p \simeq 10^{-8}$ . We can
obtain the number of events for $N_{bg}$ background stars with $T$
exposure time as
\begin{equation}
N_p = \frac{\pi}{2} \frac{T N_{bg}}{<t_E>}\times \tau_p,
\end{equation}
where the average Einstein crossing time for the Galactic Bulge
events is about $<t_E> = 28~days$. In Table (\ref{tab2}) we provide
the numerical values for the optical depth and number of events for
each photometric cases.

\section{Conclusion}
\label{conc} In this work we examined the possibility of hot Jupiter
detection through caustic crossing of a binary lens by a planet as
the source object. The effect of this caustic crossing due to the
small size of the planet is like an illumination on the microlensing
light curve of the parent star. Taking the flux of the parent star
as the background light and the illumination of the planet as the
signal, we studied the physical characteristics of the planet as the
orbital size, the atmospheric property and the size of the planet
from one hand and the characteristics of the parent star and lens
from the other hand on the observability of planet.

In the next step, we did a Monte-Carlo simulation to obtain the
detection efficiency of the planet with this method. We
showed that takeing just geometrical caustic crossing of the
planets, $36\%$ of population can be illuminated. However in reality
due to the photometric error the peak generated by the planet may
not be detected. We used three different photometric precision in
the observation and obtained the detection efficiency,
assuming that we use a network of the telescopes to have enough
sampling of data much less than one hour, a typical time of the
caustic crossing of the planet. We showed that for longer pass-bands
the detection efficiency is increased due to the more relative
emissivity of planet compare to the parent star.

Finally we estimate the number of the Hot Jupiters that can be observed
with this method towards the Galactic bulge. With a ten years monitoring
of $10^7$ stars towards the Galactic Bulge, we can detect in the
order of $10$ Hot Jupiter with this method. This observation may be done by the
next generation of the microlensing surveys towards the Galactic
Bulge.

 \end{document}